%% file: main.tex
\documentclass[sigconf,natbib=true,square,numbers]{acmart}
\usepackage{multirow}
\usepackage{makecell}
\usepackage{graphicx}
\usepackage{geometry}
\usepackage{array}
\usepackage{xspace}
\usepackage{listings}
\usepackage{xcolor}
\usepackage{appendix}
\usepackage{caption}
\usepackage{booktabs}
\usepackage{amsmath}
\usepackage{float}
\usepackage{url}

\usepackage{algorithm}
\usepackage{algpseudocode}

\lstdefinestyle{promptstyle}{
    backgroundcolor=\color{gray!10},
    basicstyle=\ttfamily\fontsize{7}{9}\selectfont,
    frame=single,
    breaklines=true
}

\input{macros}
\AtBeginDocument{%
  \providecommand\BibTeX{{%
    \normalfont B\kern-0.5em{\scshape i\kern-0.25em b}\kern-0.8em\TeX}}}

\input{category}
\makeatletter

\begin{document}

\title{On the Merits of LLM-Based Corpus Enrichment}
\settopmatter{printacmref=false, printfolios=false}


\author{Gal Zur}
\email{gal.zur@campus.technion.ac.il}
\affiliation{%
  \institution{Technion}
  \city{Haifa}
  \country{Israel}
}

\author{Tommy Mordo}
\email{tommymordo@technion.ac.il}
\affiliation{
  \institution{Technion}
  \city{Haifa}
  \country{Israel}
}

\author{Moshe Tennenholtz}
\email{moshet@technion.ac.il}
\affiliation{%
  \institution{Technion}
  \city{Haifa}
  \country{Israel}
}

\author{Oren Kurland}
\email{kurland@technion.ac.il}
\affiliation{%
  \institution{Technion}
  \city{Haifa}
  \country{Israel}
}

\renewcommand{\shortauthors}{Zur et al.}

\input{abstract}
\maketitle
\input{introduction}

\input{related}
\input{generation}
\input{direct-search}
\input{rag}
\input{attribution}
\input{conc}
\balance
\bibliographystyle{ACM-Reference-Format}
\input{main.bbl}

\end{document}

%% file: macros.tex
\newcommand{\omt}[1]{}

\newcommand{\firstmention}[1]{{\bf #1}}

\newcommand{\myparagraph}[1]{\vspace{0.3\baselineskip}\noindent{\textbf{#1}}.~}

\newcommand{\query}{q}
\newcommand{\answer}{a}

\newcommand{\corpus}{\mathcal{C}}
\newcommand{\subcorpus}{\mathcal{C'}}
\newcommand{\encorpus}{\mathcal{C}\cup{\{\dgen |q \in Q\}}}
\newcommand{\corpuse}{\mathcal{C}_e}

\newcommand{\noen}{NoEnrich}

\newcommand{\dlf}{DL21}
\newcommand{\dls}{DL22}
\newcommand{\ms}{MS-MARCO V2}

\newcommand{\gpt}{GPT-4}
\newcommand{\la}{Llama2}
\newcommand{\lb}{Llama3}

\newcommand{\gpttable}{G4}
\newcommand{\latable}{L2}
\newcommand{\lbtable}{L3}

\newcommand{\okapi}{Okapi}
\newcommand{\ef}{E5}
\newcommand{\cont}{Contriever}

\newcommand{\okapitable}{Ok}
\newcommand{\eftable}{E5}
\newcommand{\conttable}{Co}

\newcommand{\dgen}{d_{q}^{g}}
\newcommand{\zs}{ZS}
\newcommand{\ods}{DM}
\newcommand{\tds}{2DS}
\newcommand{\trds}{3DS}
\newcommand{\tdsr}{2DSR}

\newcommand{\nrag}{no-RAG}
\newcommand{\ragne}{RAG-no-Enrich}
\newcommand{\rage}{RAG-Enrich}

\newcommand{\datt}{d_{q}^{a}}

\newcommand{\aisp}{Acc}
\newcommand{\maisp}{Acc-NoGen}
\newcommand{\okr}{BM25}
\newcommand{\oknr}{BM25+NLI}



%% file: category.tex
\ccsdesc[500]{Information systems}
\ccsdesc[500]{Information systems~Information retrieval}
\ccsdesc[500]{Information systems~Search engine architectures and scalability}
\ccsdesc[500]{Information systems~Adversarial retrieval}

\keywords{LLM, corpus enrichment, RAG}

%% file: abstract.tex
\begin{abstract}
  Generative AI (genAI) technologies --- specifically, large language
  models (LLMs) --- and search have evolving relations. We argue for a
  novel perspective: using genAI to enrich a document corpus so as to
  improve query-based retrieval effectiveness. The enrichment is based
  on modifying existing documents or generating new ones.
  As an empirical proof of concept, we use LLMs to generate
  documents relevant to a topic which are more retrievable than existing ones. In addition, we demonstrate the potential merits of
  using corpus enrichment for retrieval augmented generation (RAG) and answer attribution in question answering.
  
\end{abstract}

%% file: introduction.tex
\section{Introduction}
\label{sec:intro}
Generative AI (genAI) technologies are the basis for the potential
shift from 10-blue-links search engines to (conversational) question
answering (QA) systems (e.g., \cite{perplexity,you}). Nevertheless, there
is an in-depth reliance of QA systems on search engines. A case in
point, retrieval augmented generation (RAG) is among the most
effective approaches for providing fresh answers which are grounded
\cite{lewis2020retrieval,izacard2022few,Gao+al:24a}. Furthermore,
attribution in QA, namely presenting supportive evidence for answers,
often relies on search \cite{Bohnet+al:23a}. On the other hand, there
is a growing body of work on using genAI technologies for search;
e.g., using LLMs for ranking 
\cite{Liang+al:23a,Qin+al:24a,Zhuang+al:24a}, inducing relevance
judgments \cite{Faggioli+al:23a,Clarke+Dietz:24a,Upadhyay+al:24a},
and reformulating queries \cite{Jagerman+al:23a}.

In this paper we argue for a novel perspective about how genAI can
improve query-based retrieval over a given corpus and provide an
initial empirical support to the merits of this
perspective. Specifically, we advocate the use of large language
models (LLMs) to modify documents in a corpus and to create new
ones. That is, the goal is to enrich the corpus in an offline manner so as to improve the effectiveness of query-based retrieval.

Consider the fundamental situation where retrieval of documents
relevant to an information need (topic) is a difficult challenge. The
challenge can be identified, for example, by applying query log
analysis or by using query performance predictors
\cite{Yom-Tov+al:05a}. The difficulty could be due to the user's
choice of a query to represent the information need, the retrieval
function applied or other reasons. For example, for lexical retrieval
methods (e.g., Okapi BM25), the vocabulary mismatch between the query
and relevant documents is a challenge which drove forward work on
automatic query expansion
\cite{Buckley+al:94a,Carpineto:Romano:12a}. Now, generating additional
relevant documents which are
``retrievable'' \cite{Azzopardi:15a} for various ranking functions
and/or selection of a query to represent the information need \cite{Bailey+al:16a} can potentially aleviate
the retrieval challenge. For example, to
address the vocabulary mismatch between a query and relevant
documents, we can use an LLM to slightly modify existing relevant
documents taking care to (i) include query terms in the modified document, (ii) not hurt
discourse and overall document quality, and (iii) maintain
faithfulness to the content of the original relevant document. 

An approach similar to that just described can be applied to address
the significant variance in the retrieval performance of using
different queries to represent the same information need
\cite{Bailey+al:16a,Zendel+al:19a}, a.k.a., query variants
\cite{Bailey+al:16a}. That is, LLMs can be used to create versions of
existing relevant documents that contain different query variants. As
recently noted, the corpus has major effect on the difficulty of a
query, that is, the resultant retrieval performance
\cite{Culpepper+al:22a}. Here we propose to enrich an existing corpus
so as to reduce variance with respect to query variants.

Fairness of exposure of documents' authors in ranked lists
\cite{Castillo+al:18a} is an additional aspect that can benefit from
corpus enrichment. While there is a large body of work on ranking
functions that promote fairness \cite{Ekstran+al:22a}, having authors
generate versions of their documents which are more retrievable can potentially have major positive impact. In a related vein, recent work shows how LLMs can be used to modify documents so as to improve their future ranking in competitive search settings \cite{Bardas+al:25a}; i.e., settings where authors opt to have their documents highly ranked for queries of interest.

Summarization and distillation of relevant information in newly created
documents can also benefit the search ecosystem. Summarizing information relevant to
a topic in a single document can alleviate the need for evolved
diversification algorithms \cite{Santos+al:15a,wu_result_2024} and
ameliorate users' efforts in browsing the retrieved list and
aggregating the relevant information. On the other hand, distilling
relevant information from documents whose majority of content is not
relevant to a topic can help to improve their
retrievability\footnote{Often, the majority of content in documents
  relevant to a query is not relevant \cite{Sheetrit+al:20a}.}. 

An important aspect of (LLM-based) corpus enrichment, specifically for the goals
described above, is faithfulness to the content present in the
corpus. Indeed, a major issue with content generated using LLMs is
hallucinations \cite{Gao+al:24a}. We present novel
evaluation measures for the faithfulness of newly generated documents
 to content in (i) the documents based on which generation was performed, and
(ii) the entire corpus. Furthermore, we show that corpus
enrichment can benefit RAG, which is the main paradigm to
reducing hallucinations \cite{Gao+al:24a}.

Additional important question about corpus enrichment is ownership and
responsibility with respect to the generated content. Putting aside
important copyright issues about content generated by LLMs, we
differentiate two types of search settings. The first is a setting
with a single (centralized) owner of the corpus. A canonical example
is enterprise retrieval where the owner of all content is the
enterprise. Accordingly, the enterprise is responsible for the content
generated for corpus enrichment. The second type of search setting is
a corpus with multiple authors and no centralized
owenrship/responsibility. The Web is obviously the prominent example
of such a setting. As is the case with manually created content,
publishers using genAI to perform corpus enrichment (e.g., modifying
their documents) are responsible for the content they generate. More
importantly, the question is whether search engines should become
content generators; e.g., for improving retrievability of relevant
documents as discussed above. In this case, search engines will serve not only the traditional role of mediators, but also of content creators. We argue that this is already the case in some sense with genAI question answering systems on the Web \cite{perplexity,you} which generate content in response to questions. Still, the question of whether search engines should become content generators has far reaching implications the discussion of which is outside the scope of this paper.

Since question answering systems based on LLMs (e.g.,
\cite{perplexity,you}) respond to user requests (e.g., questions) by
generating answers which could be viewed as pseudo documents, one
might wonder about the relative advantages of the offline corpus
enrichment perspective we propose. The first is democratization of
content generation: while all stakeholders can contribute to corpus
enrichment using various genAI tools, a given QA system is
centralized. On a different note, offline document creation processes
can be more in-depth and on-going than on-the-fly response
generation. Additional observation is that online response generation
still relies on search, with RAG and attribution as noted above being
prominent examples; we show below that RAG and attribution can much benefit from
corpus enrichment. Finally, some researchers argue for the importance
of the user information seeking experience in search processes with
respect to being provided with an answer to a question
\cite{Shah+Bender:24a}; our corpus enrichment proposal supports
traditional search processes.

Below we present a first proof of concept for the merits of the
proposed perspective of corpus enrichment. We create new documents in
a corpus by prompting LLMs so as to improve the effectiveness of
query-based retrieval in three scenarios: (i) standard ad hoc
retrieval, (ii) retrieval augmented generation (RAG) in LLM-based
question answering, and (iii) attribution, via passage retrieval, in
LLM-based question answering. Our generation methods are quite simple:
they are mainly based on query-biased document
summarization/rephrasing performed by prompting LLMs. Still, we show that relevant documents generated based on existing ones, 
can be substantially more retrievable; specifically, they are ranked at positions higher
than the median rank of existing relevant documents. As a result, retrieval effectiveness is substantially improved. In addition, we demonstrate the potential in improving LLM-based question answering performance by applying corpus enrichment which benefits the retrieval stage in RAG. Finally, we show how passage retrieval used for attribution in LLM-based question answering can significantly improve if corpus enrichment is applied.

Our contributions can be summarized as follows:
\begin{itemize}
\item A novel perspective about using LLM-based corpus enrichment to improve query-based retrieval effectiveness.
\item An initial empirical study that provides support to the merits of the proposed perspective in three query-based retrieval settings: (i) ad hoc retrieval, (ii) retrieval augmented generation for question answering, and (iii) attribution for answers in question answering.
  \item Novel measures of the faithfulness of generated content to that in a corpus.
  \end{itemize}

%% file: related.tex
\section{Related Work}
Expanding a document or enriching its representation has long been
used as a means to coping with the vocabulary mismatch between queries
and relevant documents
\cite{Kurland+Lee:04a,Liu+Croft:04a,Wei+Croft:06a,Efron+al:12a,Nogueira+al:19a}. In
contrast to this line of work, we create new documents in the corpus
(and modify existing ones) using LLMs to address various issues.

Documents generated using LLMs were used to train ranking functions
\cite{Askari+al:23a}. Our goal, in contrast, is to have the generated
documents retrieved.

In competitive search settings \cite{Kurland+Tennenholtz:22a}, document authors modify their documents so as to improve their future ranking by an undisclosed ranking functions. There is a line of work on automatic document modification to this end \cite{Goren+al:20a,Bardas+al:25a} including LLM-based techniques \cite{Bardas+al:25a}. As argued above, such methods for corpus enrichment can potentially be used to potentially improve the fairness in exposure of documents' authors \cite{Castillo+al:18a,Ekstran+al:22a}.

In the emerging generative retrieval paradigm \cite{li2025matchinggenerationsurveygenerative}, given a query, IDs of presumably relevant documents in the corpus are generated. In contrast to our corpus enrichment approach, no new documents are generated to augment the corpus.

%% file: generation.tex
\section{Document generation methods} \label{sec:gen}

We present methods to generate documents so as to enrich a corpus. The goal is to improve the effectiveness of query-based retrieval. In Sections \ref{sec:direct}, \ref{sec:rag} and \ref{sec:att} we show how the enrichment can help to improve the effectiveness of standard ad hoc document retrieval, retrieval augmented generation for question answering and attribution in question answering. Our methods are agnostic to the retrieval settings and are based on prompting an LLM. The prompts include a given query or question (henceforth also referred to as a query; denoted $\query$) and possibly one or more existing relevant document(s) from a corpus $\corpus$ so as to generate a new document. The query $\query$ is selected from a set of questions $Q$. We define the following prompts using which the LLM will generate a new document $\dgen$:


\begin{itemize}
\item \firstmention{Zero-Shot Content} (Figure \ref{zs_prompt}; denoted \textbf{\textit{ZS}}). In this method, the LLM is prompted to generate a document containing relevant content based on the information need inferred from a query. This zero-shot approach leverages the model’s pre-existing knowledge \cite{zeroshot} based on the presumed information need underlying the query.

\item \firstmention{Document Modification} (Figure \ref{ods_prompt}; denoted \textbf{\textit{\ods{}}}). The context in this approach includes the query ($\query$) and a single document. The prompt instructs the LLM to rewrite the given document without introducing any information not explicitly expressed in the original text; note that the resultant document is a query-biased rephrasing.

\item \firstmention{Two-Document Summary} (Figure \ref{summary_prompt}; denoted \textbf{\textit{\tds}}). This approach extends the \ods{} method by incorporating a second document, providing the LLM with additional context for generating the summary; the generated document is essentially  a query-biased summary.  

\item \firstmention{Three-Document Summary} (Figure \ref{summary_prompt}; denoted \textbf{\textit{\trds}}). This method is similar to the \textit{\tds{}} method, but it includes in the context an additional third document, allowing the LLM to generate a potentially more comprehensive summary with a broader contextual understanding of the underlying information need represented by the query.
\end{itemize}

\begin{figure}[t]
\centering
\input{Images/zero-shot}
\caption{The prompt for zero-shot document generation (\zs{}).}
\label{zs_prompt}
\end{figure}

\begin{figure}[t]
\centering
\input{Images/ods}
\caption{The prompt for document modification (\ods{}).}
\label{ods_prompt}
\end{figure}

\begin{figure}[t]
\centering
\input{Images/summary}
\caption{The prompt to generate a query-biased summary of two or three documents (\tds{} and \trds{}).}
\label{summary_prompt}
\end{figure}


We create an enriched corpus per generation method, incorporating the generated documents. We generate one document per query; the resultant corpus is denoted $\corpuse = \encorpus$.

\myparagraph{LLMs}
We utilized \gpt{} \cite{achiam2023gpt}, \la{} \cite{touvron_llama2}, and \lb{} \cite{grattafiori_llama_2024} as the LLMs\footnote{meta-llama/Meta-Llama-3-8B-Instruct, meta-llama/Llama-2-13b-chat-hf from Hugging face repository.} for document generation, using the prompting methods described above. To minimize the likelihood of hallucinations, the temperature parameter was set to zero. When the temperature is zero, the LLM consistently selects the most probable tokens during the auto-regressive generation process.

%% file: Images/zero-shot.tex
\begin{lstlisting}[style=promptstyle]
PROMPT = "You are a content provider. Write a document that has relevant information to the need induced by a given query. Write it as a short paragraph. You must remain truthful, while also making sure your paragraph would be ranked higher than other paragraphs of the same topic. Write a short paragraph that satisfies the information need induced by the query: <query>"
\end{lstlisting}

%% file: Images/ods.tex
\begin{lstlisting}[style=promptstyle]
PROMPT = "Rewrite a given document, according to a given query. Do not add new knowledge not present in the document. It should be similar to the original, and should answer the query. Write it in a paragraph form. Rewrite according to the query: <query> <document>}"
\end{lstlisting}

%% file: Images/summary.tex
\begin{lstlisting}[style=promptstyle]
PROMPT = "Your task is abstractive summarization of given documents, according to a given query. You may only use the information given to you. Do not add knowledge not present in the documents. Write it in a paragraph form. Summarize only the relevant information as induced by the following query: <query <document 1> <document 2> <document 3 (if needed)>"
\end{lstlisting}

%% file: direct-search.tex
\section{Ad hoc Retrieval}
\label{sec:direct}
In this section, we evaluate the potential merits of corpus enrichment for ad hoc retrieval. The idea is to augment the set of documents relevant to a query by generating new relevant documents that are retrievable — that is, highly ranked by standard retrieval models. This task is particularly important given that some queries have very few relevant documents, and some of these may be difficult for retrieval systems to retrieve \cite{Castillo:18a}. A query having very few relevant documents can be identified using query log analysis or by applying query performance prediction \cite{Yom-Tov+al:05a}. Some of our generation methods need a few examples of relevant documents. We assume that these examples can be obtained using click-through data or explicit relevance feedback. Studying the merits of applying generation using pseudo relevant documents is left for future work.

We explore the impact of corpus enrichment on retrieval effectiveness for each generation method introduced in Section \ref{sec:gen}. To this end, we employ two LLMs for document generation: \lb{} \cite{grattafiori_llama_2024} and \gpt{} \cite{achiam2023gpt}. The evaluation is based on two passage retrieval datasets: TREC Deep Learning 2021 (\dlf{}) \cite{TREC_DL_2021} and TREC Deep Learning 2022 (\dls{}) \cite{TREC_DL_2022}.

\subsection{Experimental setup}
\myparagraph{Datasets}
We used the datasets \dlf{} \cite{TREC_DL_2021} and \dls{} \cite{TREC_DL_2022}. They are based on \ms{} \cite{bajaj2018msmarcohumangenerated} and serve for passage ranking tasks. The \ms{} passages (documents) \cite{bajaj2018msmarcohumangenerated} were indexed (Porter stemmed) using Pyserini \cite{Lin_etal_SIGIR2021_Pyserini}, which was then used both for collecting corpus statistics and for computing ranking effectiveness measures. The \dlf{} and \dls{} datasets contain 53 and 76 queries, respectively. Document relevance judgments in both datasets are graded on a scale from zero to three. Following the \dlf{} guidelines \cite{TREC_DL_2021}, we consider documents with a relevance grade of at least two as relevant.

\myparagraph{Rankers}
We employ three unsupervised ranking functions that were not trained on \ms{}: two dense retrieval methods and one sparse retrieval method. The dense retrieval methods are \ef{} \cite{wang_text_2024} and \cont{}\footnote{intfloat/e5-large-unsupervised and facebook/contriever from the Hugging Face repository.} \cite{izacard_unsupervised_2022}, and the sparse retrieval method is \okapi{} BM25 \cite{okapi2}. For the dense retrieval methods, the retrieval score is the cosine between the query and document embedding vectors.

\myparagraph{Document generation}  
Several generation methods introduced in Section \ref{sec:gen} (specifically, \ods{}, \tds{}, \trds{}) require relevant documents for a given query. To select these documents, we apply a retrieval-based selection process.

For each query, we retrieve 1000 documents using \okapi{} \cite{craswell_trec_dl_2019}. These documents are divided into ordered groups of ten based on their retrieval rank. From each group that contains at least one relevant document, we randomly select one relevant document; existing relevance judgments are used \cite{bajaj2018msmarcohumangenerated}. If the number of selected documents from the ranked groups is less than three (for the \trds{} generation method), the remaining relevant documents are randomly sampled from the full set of relevant documents. To study the merit of using relevant documents rather than using documents that are selected randomly, we introduce an additional generation method, denoted \textbf{\tdsr{}}. This method uses two documents as a context to the LLM, one relevant document and one document that is selected randomly from the top-1000 documents retrieved by \okapi{}, excluding the relevant documents.
Each generation method is then applied to generate one document per query for each LLM. Relevance judgments of the generated documents are obtained as described below.

\myparagraph{Ranking effectiveness}
We begin by generating one document per query for each LLM and generation method, resulting in ten corpora based on \ms{} passages (documents) — one for each combination of LLM and generation method. Next, we use \okapi{} to retrieve 10,000 documents for each query; at least one relevant document per query is ranked within the 10,000 documents. The ranking per query is induced over the corpus after documents generated for all the other queries were removed. Then, the retrieved set of 10,000 documents is re-ranked using \cont{} and \ef{}. We report the median rank of the generated document (over queries; denoted \textbf{MG}) for the three ranking functions (\okapi{}, \cont{} and \ef{}). If a generated document is not among the 10,000 retrieved documents, it is assigned a rank of 20,000. 

For the corpus before enrichment (denoted \textbf{\noen{}}), we report two statistics: (i) the median rank of relevant documents per query among the top 10,000 retrieved documents; then we compute the median over the queries (denoted \textbf{ME}), and (ii) the median rank of the highest-ranked relevant document over the queries (denoted \textbf{HR}).

For all the corpora, we also report NDCG@10, NDCG@100, and MAP@100, following the standard in \dlf{} and \dls{}\footnote{In \dlf{} and \dls{}, NCG@100 was originally computed; however, for implementation reasons, we computed NDCG@100, which is correlated with NCG@100.}. To compute NDCG@10, NDCG@100, and MAP@100 for \ef{} and \cont{} we re-rank only 100 documents retrieved by \okapi{}, following the standard in \dlf{} and \dls{} \cite{TREC_DL_2021,TREC_DL_2022}.

\myparagraph{Relevance judgements}  
Each generated document was judged for relevance to the corresponding query by three English-speaking crowd workers on the Connect platform via CloudResearch \cite{noauthor_introducing_2024}. The judgement guidelines followed those for \dlf{} \cite{TREC_DL_2021} and \dls{} \cite{TREC_DL_2022}. Each document was assigned one of the following relevance grades: (i) 0: not relevant to the query, (ii) 1: related to the query, (iii) 2: highly relevant to the query, and (iv) 3: perfectly relevant to the query. The final relevance grade of each document was determined as the median score across the three annotators. The inter-annotator agreement rates for the generated documents over the generation methods, measured using free-marginal multi-rater Kappa, were $72\%-90\%$ and $65\%-82\%$ for the generated documents in the \dlf{} dataset for \gpt{} and \lb{}, respectively. In the \dls{} dataset the Kappa rates ranged $44\%-80\%$ for \gpt{} and $46\%-77\%$ for \lb{}. For both datasets, the highest inter-annotator agreement rate was observed for the \zs{} generation method.

$88.8\%-100\%$ and $81.1\%-100\%$ of the generated documents were relevant in the \dlf{} dataset for \gpt{} and \lb{}, respectively. For the \dls{} dataset, $72.4\%-96.1\%$ and $78.9\%-94.7\%$ of the generated documents were judged as relevant for \gpt{} and \lb{}, respectively. These findings attest to the fact that it is possible to generate documents which with relatively high likelihood are relevant using the presented generation methods (as described in Section \ref{sec:gen}). Interestingly, the \zs{} generation method produced the highest number of relevant documents. The full details are presented in Table \ref{tab:ds:rel}.

\begin{table}[t]
\caption{The percentage of relevant documents for \dlf{} and \dls{} for the LLMs and generation methods.}
\label{tab:ds:rel}
\centering
\input{Tables/rel}
\end{table}

\subsection{Faithfulness}
Using LLMs to generate documents raises concerns about hallucinations \cite{shuster2021retrieval}, potentially resulting in false and misleading information. We therefore measure the extent to which the content in the generated document is ``faithful`` to the information in the corpus. To this end, we use the NLI (Natural Language Inference) model presented by Honovich et al. \cite{honovich-etal-2022-true-evaluating}. Inspired by previous work on measuring the level of attribution \cite{Bohnet+al:23a, Gao_RARRRA}, we use the model to estimate whether one document is entailed by another while preserving factual consistency. The model outputs a score in the range $[0,1]$, where higher values indicate stronger factual alignment between the texts.

To apply the NLI approach in our setting, we use a \\ query-dependent sample of the corpus, denoted $\subcorpus \subseteq \corpus$. Specifically, we adopt a greedy (approximation) algorithm, inspired by the hill climbing algorithm \cite{nemhauser_hill_climbing} that iteratively finds up to $k$ documents in $\subcorpus$ for each sentence in $\dgen$, the generated document, ensuring that together they entail the sentence; entailment is determined using a threshold of 0.5 \cite{Bohnet+al:23a} on the NLI. The set of these (up to) $k$ documents is referred to as the Knowledge Base (KB) of the sentence. If the greedy algorithm fails to find $k$ such documents, it returns an empty set; when the returned set is empty, the entailment score is set to 0 for that sentence. The full details of the method of constructing a knowledge base for a sentence is presented in Algorithm \ref{algorithm:faithfulness}.
To segment documents into sentences, we use the boundary detection algorithm by Kiss et al. \cite{kiss-strunk-2006-unsupervised}, implemented in the NLTK platform \cite{loper_nltk_2002}. The resultant \firstmention{faithfulness score} of a generated document $\dgen$ is defined as the percentage of entailed sentences in $\dgen$; note that different sentences may be entailed by different sets of $k$ documents.

\begin{algorithm}[H]
\caption{Knowledge Base Construction Using Hill Climbing}
\label{algorithm:faithfulness}
\input{Algorithms/faithfulness}
\end{algorithm}

We use two different samples ($\subcorpus$) of the corpus to determine the faithfulness of the generated documents: (i) the set of relevant documents, and (ii) the union of the relevant documents and the 1000 most highly ranked documents by \okapi{}. We use all relevant documents as some might not be among the 1000 most highly ranked.

As a reference for measuring the faithfulness of the generated document, we also measure the average faithfulness of the existing set of relevant documents; specifically, we randomly selected five relevant documents, using the same procedure as that used to select relevant document for generating new documents. Then, we compute their average faithfulness score, with respect to both samples described above; we excluded the set of five relevant documents from the sample $\subcorpus$ served to entail that set.

\subsection{Results}

We present the ranking effectiveness and faithfulness results\footnote{Statistically significant differences are determined using the two-tailed paired permutation test
with 100,000 random permutations, using a 95\% confidence level.} for the \dlf{} and \dls{} datasets in Tables \ref{tab:trec_dl21_ranking} and \ref{tab:ds_faith}.

We start by analyzing the ranking effectiveness results for the corpus without enrichment and with enrichment. 
All the measures are reported for every generation method and LLM used to generate the documents. We report the results (Table \ref{tab:trec_dl21_ranking}) for two LLMs (\lb{} and \gpt{}), the three ranking functions (\okapi{}, \ef{}, and \cont{}), and the five generation methods from Section \ref{sec:gen}.  

We see in Table \ref{tab:trec_dl21_ranking} that the median rank of the generated documents (MG) is statistically significantly lower than the median rank of the existing relevant documents (ME) across all ranking functions and generation methods. This finding indicates that LLM-generated documents have the potential to be highly ranked. Note that for the generated documents, we report the median over queries, where for each query there is a single document that was generated. The comparison is with the baselines, for which we compute the median over queries of the median rank of the (existing) relevant documents per query, or with the median (over queries) of the highest-ranked relevant document per query. While the numerical differences may appear substantial, they can be attributed to the fact that we compare statistics of a measure of a single generated document to a set of documents retrieved per query.

As we can see in Table \ref{tab:trec_dl21_ranking}, among the ranking functions, the highest ranks were observed for \ef{}, which is aligned with prior reports \cite{wang_text_2024,izacard_unsupervised_2022}. Specifically, for \dlf{}, the median rank of relevant documents without enrichment was $309$, $146$, and $246$ for \okapi{}, \ef{}, and \cont{}, respectively. The highest (over generation methods and LLMs) median rank of the generated documents was $9$, $1$, and $39$ for \okapi{}, \ef{}, and \cont{}, respectively for \dlf{} and $4.5$, $1$ and $46.5$, respectively for \dls{}. Interestingly, the median rank of generated documents was higher for \okapi{} than for \cont{}, whereas the median rank of existing relevant documents was higher for \cont{} than for \okapi{}. Additionally, we observe in Table \ref{tab:trec_dl21_ranking} that the effectiveness measures computed for \ef{} were higher than those for \okapi{} and \cont{}. This finding is in accordance with our finding that both relevant and generated documents receive higher ranks with \ef{} than with other ranking functions.


Overall, we can see in Table \ref{tab:trec_dl21_ranking} that ranking effectiveness for the enriched corpus is better than that for the original corpus, for all generation methods and ranking functions; some of the improvements are statistically significant. This result can be attributed to the fact that (i) the generated documents are ranked at higher positions than the median rank of the existing relevant documents and (ii) most of them are relevant as was show in Table \ref{tab:ds:rel}. For instance, the NDCG@10 for \okapi{} is at most $0.481$ for \zs{} with \lb{} while the baseline NDCG@10 (without corpus enrichment) is $0.446$. For \ef{} and \cont{}, the \trds{} generation method performed slightly better than the other generation methods; this finding aligns with the fact that the \trds{} generation method uses more relevant documents for generation compared to the other generation methods.

\newcommand{\medDiff}{^{m}}
\newcommand{\topDiff}{_{h}}
\newcommand{\retDiff}{^{*}}
\begin{table}[t]
\caption{Retrieval effectiveness of three ranking functions (\okapi{} and \cont{} are abbreviated as \textbf{Ok} and \textbf{Co}, respectively) for \dlf{} and \dls{} when using the five document generation methods and two LLMs (\lb{} and \gpt{}, denoted \textbf{L3} and \textbf{G4}, respectively). We present the median rank position of the generated documents (\textbf{MG}), \textbf{NDCG@10} (N@10), \textbf{NDCG@100} (N@100), and \textbf{MAP@100} (MAP). We present for reference the median (over queries) of the median rank of the existing relevant documents per query (\textbf{ME}) and the median (over queries) rank position of the highest ranked relevant document (\textbf{HR}) in the corpus without enrichment. Statistically significant differences of \textbf{MG} with \textbf{ME} and \textbf{HR} are marked with "$m$" and "$h$", respectively. Statistically significant differences of the evaluation measure results between the enriched corpus (five generation methods) and without enrichment (\noen) are marked with a "$*$". The best ranking performance in a column (per dataset) is boldfaced.}
\label{tab:trec_dl21_ranking}
\centering
\input{Tables/DS_table_dl21}
\end{table}

\myparagraph{Faithfulness results}
We report in Table \ref{tab:ds_faith} the average (over generated documents) faithfulness scores with respect to (i) the set of existing relevant documents and (ii) an approximation sample of the corpus (which includes the relevant documents). Algorithm \ref{algorithm:faithfulness} depends on the number of documents $k$ used to entail each sentence in the generated document. We report the faithfulness scores for $k=1$ and $k=5$, as the results for higher values of $k$ tend to saturate.

We observe in Table \ref{tab:ds_faith} that for $k=5$, the faithfulness scores for \tds{} and \trds{} exceed $90\%$ across both LLMs and datasets. In contrast, the \zs{} generation method exhibits significantly lower faithfulness. This difference can be attributed to the underlying mechanisms of the generation methods: in \tds{} and \trds{}, documents from the corpus are provided to the LLM as auxiliary information to guide document generation. In contrast, \zs{} which is based on zero-shot generation lacks such auxiliary information, leading to the creation of documents that may contain content drifting from that in the corpus.

Additionally, we can see in Table \ref{tab:ds_faith} that the faithfulness score of a random sample of five relevant documents (the baseline) is lower than that of \tds{} and \trds{}. This finding suggests that the documents generated using \tds{} and \trds{} are inherently more faithful, as they are conditioned on relevant corpus documents provided as context. 


\newcommand{\faithDiff}{^{*}}
\begin{table}[t]
\caption{The level of faithfulness of the generated documents. We report the averaged (over generated documents) faithfulness scores for the five generation methods, two LLMs (\lb{} and \gpt{}, denoted L3 and G4, respectively), and two values of $k$ ($k=1$ and $k=5$); recall that $k$ is the number of documents from the original corpus $\corpus$ used to entail a generated document. We report the faithfulness scores for two corpus samples : (i) the set of relevant documents for the query (denoted \textbf{Rel}), and (ii) the union of the relevant documents for the query and the 1000 documents retrieved by \okapi{} (denoted \textbf{Corpus}). As a baseline, we report the average faithfulness score of the existing relevant documents (\textbf{RD}). Statistically significant differences with the \textbf{RD} baseline are marked with "$*$". The best result in a column (per dataset) is boldfaced.}
\label{tab:ds_faith}
\centering
\input{Tables/DS_faith_dl21}
\end{table}

%% file: Tables/rel.tex
\begin{scriptsize}
\setlength{\tabcolsep}{4pt}
\begin{tabular}{ll*{5}{c}}
    \toprule
    & & \textbf{\zs{}} & \textbf{\ods{}} & \textbf{\tds{}} & \textbf{\tdsr{}} & \textbf{\trds{}} \\
    \midrule
    \multicolumn{7}{c}{\textbf{DL 2021}} \\
    \toprule
    \textbf{\gpt{}} & & 100.0 & 90.6 & 96.2 & 88.8 & 98.1 \\
    \textbf{\lb{}} & & 100.0 & 94.3 & 98.1 & 81.1 & 96.2 \\
    \midrule
    \multicolumn{7}{c}{\textbf{DL 2022}} \\
    \toprule
    \textbf{\gpt{}} & & 96.1 & 72.4 & 89.5 & 73.7 & 93.4 \\
    \textbf{\lb{}} & & 94.7 & 85.5 & 78.9 & 78.9 & 88.2 \\
    \bottomrule
\end{tabular}
\end{scriptsize}

%% file: Algorithms/faithfulness.tex
\begin{algorithmic}[1]
\Require A sentence \( s \), an NLI model \( m \), an integer number \( k \), and a subset of the corpus \( \subcorpus \subset \corpus \)
\Ensure \( KB_k \) or \( \emptyset \)
\State \( KB_0 \gets \emptyset \)
\For{\( i \gets 1 \) to \( k \)}
    \State Pick some \( d \in \arg\max_{d \in \mathcal{C}'\setminus KB_{i-1}} NLI(m, KB_{i-1} \cup \{d\}, s) \)
    \State \( KB_i \gets KB_{i-1} \cup \{d\} \)
    \If{\( NLI(m, KB_i, s) \geq 0.5 \)}
        \Return \( KB_i \)
    \EndIf
\EndFor \\
\Return \( \emptyset \)
\end{algorithmic}

%% file: Tables/DS_table_dl21.tex
\begin{scriptsize}
\setlength{\tabcolsep}{1pt}
\begin{tabular}{ll*{12}{c}}
    \toprule
     &  
    & \multicolumn{3}{c}{\textbf{MG}} 
    & \multicolumn{3}{c}{\textbf{N@10}} 
    & \multicolumn{3}{c}{\textbf{N@100}} 
    & \multicolumn{3}{c}{\textbf{MAP}} \\
    \cmidrule(lr){3-5} \cmidrule(lr){6-8} \cmidrule(lr){9-11} \cmidrule(lr){12-14}
     & 
     & \textbf{\okapitable} & \textbf{\eftable} & \textbf{\conttable} 
     & \textbf{\okapitable} & \textbf{\eftable} & \textbf{\conttable} 
     & \textbf{\okapitable} & \textbf{\eftable} & \textbf{\conttable} 
     & \textbf{\okapitable} & \textbf{\eftable} & \textbf{\conttable} \\
    \bottomrule
    \multicolumn{14}{c}{ \textbf{\dlf{}}} \\
    \toprule
    \textbf{\noen} 
        & 
        & \makecell[c]{\textbf{ME}: 309 \\ \textbf{HR}: 2}
        & \makecell[c]{\textbf{ME}: 146 \\ \textbf{HR}: 1}
        & \makecell[c]{\textbf{ME}: 246 \\ \textbf{HR}: 3}
        & \makecell[c]{$.446$} & \makecell[c]{$.526$} & \makecell[c]{$.448$}
        & \makecell[c]{$.391$} & \makecell[c]{$.421$} & \makecell[c]{$.390$}
        & \makecell[c]{$.136$} & \makecell[c]{$.173$} & \makecell[c]{$.126$} \\
    \midrule
     \textbf{\zs} 
        & \makecell[c]{\lbtable \\ \gpttable}
        & \makecell[c]{$9\topDiff\medDiff$ \\ $18\topDiff\medDiff$}
        & \makecell[c]{$3\medDiff$ \\ $1\medDiff$}
        & \makecell[c]{$289\topDiff$ \\ $109\topDiff\medDiff$}
        & \makecell[c]{$.481$ \\ $.470$}
        & \makecell[c]{$.580$ \\ $.581$}
        & \makecell[c]{$.447$ \\ $.458$}
        & \makecell[c]{$.408$ \\ $.406$}
        & \makecell[c]{$.445$ \\ $.446$}
        & \makecell[c]{$.397$ \\ $.400$}
        & \makecell[c]{$.150$ \\ $.150$}
        & \makecell[c]{$.200$ \\ $\mathbf{.203}$}
        & \makecell[c]{$.134$ \\ $.136$} \vspace{1mm}\\
    
     \textbf{\ods} 
     & \makecell[c]{\lbtable \\ \gpttable}
        & \makecell[c]{$9\topDiff\medDiff$ \\ $17\topDiff\medDiff$}
        & \makecell[c]{$5\topDiff\medDiff$ \\ $7\topDiff\medDiff$}
        & \makecell[c]{$50\topDiff\medDiff$ \\ $69\topDiff\medDiff$}
        & \makecell[c]{$.482$ \\ $.472$}
        & \makecell[c]{$.565$ \\ $.561$}
        & \makecell[c]{$\mathbf{.465}$ \\ $.461$}
        & \makecell[c]{$.409$ \\ $.404$}
        & \makecell[c]{$.440$ \\ $.438$}
        & \makecell[c]{$\mathbf{.406}$ \\ $.400$}
        & \makecell[c]{$.151$ \\ $.145$}
        & \makecell[c]{$.198$ \\ $.193$}
        & \makecell[c]{$\mathbf{.142}$ \\ $.136$} \vspace{1mm}\\ 
    \textbf{\tds} 
    & \makecell[c]{\lbtable \\ \gpttable}
        & \makecell[c]{$10\topDiff\medDiff$ \\ $17\topDiff\medDiff$}
        & \makecell[c]{$2\medDiff$ \\ $3\medDiff$}
        & \makecell[c]{$72\topDiff\medDiff$ \\ $39\topDiff\medDiff$}
        & \makecell[c]{$.478$ \\ $.471$}
        & \makecell[c]{$.574$ \\ $.575$}
        & \makecell[c]{$.464$ \\ $.460$}
        & \makecell[c]{$.408$ \\ $.407$}
        & \makecell[c]{$.443$ \\ $.443$}
        & \makecell[c]{$.404$ \\ $.400$}
        & \makecell[c]{$.151$ \\ $.148$}
        & \makecell[c]{$.199$ \\ $.196$}
        & \makecell[c]{$.138$ \\ $.137$} \vspace{1mm}\\
     \textbf{\tdsr} 
     & \makecell[c]{\lbtable \\ \gpttable}
        & \makecell[c]{$5\medDiff$ \\ $6\medDiff$}
        & \makecell[c]{$3\medDiff$ \\ $4\topDiff\medDiff$}
        & \makecell[c]{$52\topDiff\medDiff$ \\ $170\topDiff$}
        & \makecell[c]{$.482$ \\ $\mathbf{.485}$}
        & \makecell[c]{$.562$ \\ $.573$}
        & \makecell[c]{$\mathbf{.465}$ \\ $.450$}
        & \makecell[c]{$.410$ \\ $\mathbf{.412}$}
        & \makecell[c]{$.438$ \\ $.442$}
        & \makecell[c]{$.403$ \\ $.400$}
        & \makecell[c]{$.150$ \\ $\mathbf{.156}$}
        & \makecell[c]{$.189$ \\ $.195$}
        & \makecell[c]{$.135$ \\ $.132$} \vspace{1mm} \\
      \textbf{\trds} & \makecell[c]{\lbtable \\ \gpttable}
        & \makecell[c]{$9\topDiff\medDiff$ \\ $14\topDiff\medDiff$}
        & \makecell[c]{$4\topDiff\medDiff$ \\ $3\medDiff$}
        & \makecell[c]{$100\topDiff\medDiff$ \\ $92\topDiff\medDiff$}
        & \makecell[c]{$.481$ \\ $.470$}
        & \makecell[c]{$.571$ \\ $\mathbf{.591}$}
        & \makecell[c]{$.455$ \\ $.464$}
        & \makecell[c]{$.409$ \\ $.407$}
        & \makecell[c]{$.442$ \\ $\mathbf{.450}$}
        & \makecell[c]{$.400$ \\ $\mathbf{.406}$}
        & \makecell[c]{$.150$ \\ $.147$}
        & \makecell[c]{$.195$ \\ $.201$}
        & \makecell[c]{$.136$ \\ $.141$} \\
    \bottomrule
    \multicolumn{14}{c}{ \textbf{\dls{}}} \\
    \toprule
    \textbf{\noen}
        & 
        & \makecell[c]{\textbf{ME}: 664.25 \\ \textbf{HR}: 7} 
        & \makecell[c]{\textbf{ME}: 175.25 \\ \textbf{HR}: 3} 
        & \makecell[c]{\textbf{ME}: 393.5 \\ \textbf{HR}: 5.5} 
        & \textbf{$.269$} & \textbf{$.377$} & \textbf{$.293$}
        & \textbf{$.213$} & \textbf{$.247$} & \textbf{$.224$}
        & \textbf{$.033$} & \textbf{$.059$} & \textbf{$.044$}
         \\
    \midrule
    \textbf{\zs}
        & \makecell[c]{\lbtable \\ \gpttable}
        & \makecell[c]{$25\medDiff\topDiff$ \\ $56.5\medDiff\topDiff$} & \makecell[c]{$1\medDiff\topDiff$ \\ $1\medDiff\topDiff$} & \makecell[c]{$171\medDiff\topDiff$ \\ $82\medDiff\topDiff$}
        & \makecell[c]{$.314$ \\ $.298$} & \makecell[c]{$\mathbf{.452}\retDiff$ \\ $.448\retDiff$} & \makecell[c]{$.310$ \\ $.318$}
        & \makecell[c]{$.236$ \\ $.229$} & \makecell[c]{$\mathbf{.277}$ \\ $.275$} & \makecell[c]{$.236$ \\ $\mathbf{.240}$}
        & \makecell[c]{$\mathbf{.047}$ \\ $.042$} & \makecell[c]{$\mathbf{.079}$ \\ $\mathbf{.079}$} & \makecell[c]{$.050$ \\ $.051$}
        \vspace{1mm} \\
    \textbf{\ods}
        & \makecell[c]{\lbtable \\ \gpttable}
        & \makecell[c]{$36.5\topDiff\medDiff$ \\ $56\topDiff\medDiff$} & \makecell[c]{$6\medDiff$ \\ $6.5\topDiff\medDiff$} & \makecell[c]{$59.5\topDiff\medDiff$ \\ $46.5\topDiff\medDiff$}
        & \makecell[c]{$.288$ \\ $.291$} & \makecell[c]{$.423$ \\ $.416$} & \makecell[c]{$\mathbf{.320}$ \\ $.315$}
        & \makecell[c]{$.225$ \\ $.225$} & \makecell[c]{$.265$ \\ $.263$} & \makecell[c]{$.238$ \\ $.238$}
        & \makecell[c]{$.038$ \\ $.038$} & \makecell[c]{$.070$ \\ $.070$} & \makecell[c]{$.050$ \\ $\mathbf{.053}$}
         \vspace{1mm}\\
    \textbf{\tds}
        & \makecell[c]{\lbtable \\ \gpttable}
        & \makecell[c]{$8.5\medDiff$ \\ $22\medDiff\topDiff$} & \makecell[c]{$2\medDiff$ \\ $1\medDiff$} & \makecell[c]{$46.5\medDiff\topDiff$ \\ $49.5\medDiff\topDiff$}
        & \makecell[c]{$.302$ \\ $.301$} & \makecell[c]{$.432$ \\ $.441$} & \makecell[c]{$.319$ \\ $\mathbf{.320}$}
        & \makecell[c]{$.230$ \\ $.230$} & \makecell[c]{$.269$ \\ $.271$} & \makecell[c]{$.237$ \\ $.238$}
        & \makecell[c]{$.040$ \\ $.041$} & \makecell[c]{$.069$ \\ $.075$} & \makecell[c]{$.050$ \\ $.052$}
        \vspace{1mm} \\
    \textbf{\tdsr}
        & \makecell[c]{\lbtable \\ \gpttable}
        & \makecell[c]{$4.5\medDiff$ \\ $8\medDiff$} & \makecell[c]{$1\medDiff$ \\ $2\medDiff$} & \makecell[c]{$76.5\topDiff\medDiff$ \\ $147.5\topDiff\medDiff$}
        & \makecell[c]{$.321$ \\ $.309$} & \makecell[c]{$.433$ \\ $.431$} & \makecell[c]{$.319$ \\ $.301$}
        & \makecell[c]{$.236$ \\ $.230$} & \makecell[c]{$.269$ \\ $.265$} & \makecell[c]{$.238$ \\ $.233$}
        & \makecell[c]{$.045$ \\ $.041$} & \makecell[c]{$.075$ \\ $.069$} & \makecell[c]{$.052$ \\ $.047$}
        \vspace{1mm} \\
    \textbf{\trds}
        & \makecell[c]{\lbtable \\ \gpttable}
        & \makecell[c]{$6\medDiff$ \\ $11\medDiff$} & \makecell[c]{$1\medDiff\topDiff$ \\ $1\medDiff\topDiff$} & \makecell[c]{$69.5\medDiff\topDiff$ \\ $92.5\medDiff\topDiff$}
        & \makecell[c]{$\mathbf{.326}$ \\ $.304$} & \makecell[c]{$\mathbf{.452}\retDiff$ \\ $\mathbf{.452}\retDiff$} & \makecell[c]{$.319$ \\ $.311$}
        & \makecell[c]{$\mathbf{.237}$ \\ $.230$} & \makecell[c]{$.275$ \\ $.275$} & \makecell[c]{$\mathbf{.240}$ \\ $.237$}
        & \makecell[c]{$.044$ \\ $.042$} & \makecell[c]{$.075$ \\ $.076$} & \makecell[c]{$.052$ \\ $.050$}
         \\
    \bottomrule
\end{tabular}
\end{scriptsize}

%% file: Tables/DS_faith_dl21.tex
\begin{scriptsize}
\begin{tabular}{lccccc}
    \toprule
    & & \multicolumn{2}{c}{\textbf{Rel}} & \multicolumn{2}{c}{\textbf{Corpus}}  \\
    \cmidrule(lr){3-4} \cmidrule(lr){5-6}
    & & \textbf{k=1} & \textbf{k=5} & \textbf{k=1} & \textbf{k=5} \\
    \bottomrule
    \multicolumn{6}{c}{ \textbf{\dlf{}}} \\
    \toprule
    \textbf{RD} 
        & 
        & \makecell[c]{$54.6$} 
        & \makecell[c]{$68.2$} 
        & \makecell[c]{$64.2$} 
        & \makecell[c]{$84.3$} \\
    \midrule
    \textbf{\zs{}} 
        & \makecell[c]{\lbtable \\ \gpttable} 
        & \makecell[c]{$39.4\faithDiff$ \\ $61.7$} 
        & \makecell[c]{$57.7\faithDiff$ \\ $74.6$} 
        & \makecell[c]{$46.3\faithDiff$ \\ $72.1$} 
        & \makecell[c]{$76.3\faithDiff$ \\ $92.0\faithDiff$} \vspace{1mm}\\
    \textbf{\ods{}} 
        & \makecell[c]{\lbtable \\ \gpttable} 
        & \makecell[c]{$90.8\faithDiff$ \\ $\mathbf{96.7}\faithDiff$} 
        & \makecell[c]{$97.5\faithDiff$ \\ $\mathbf{98.7}\faithDiff$} 
        & \makecell[c]{$92.7\faithDiff$ \\ $\mathbf{97.2}\faithDiff$} 
        & \makecell[c]{$98.7\faithDiff$ \\ $\mathbf{100.0}\faithDiff$} \vspace{1mm}\\
    
    \textbf{\tds{}} 
        & \makecell[c]{\lbtable \\ \gpttable} 
        & \makecell[c]{$90.8\faithDiff$ \\ $91.9\faithDiff$} 
        & \makecell[c]{$98.0\faithDiff$ \\ $97.2\faithDiff$} 
        & \makecell[c]{$91.4\faithDiff$ \\ $93.2\faithDiff$} 
        & \makecell[c]{$98.9\faithDiff$ \\ $98.9\faithDiff$} \vspace{1mm}\\

    \textbf{\tdsr{}} 
        & \makecell[c]{\lbtable \\ \gpttable} 
        & \makecell[c]{$70.6\faithDiff$ \\ $57.1$} 
        & \makecell[c]{$77.0$ \\ $61.1$} 
        & \makecell[c]{$79.5\faithDiff$ \\ $67.2$} 
        & \makecell[c]{$89.9$ \\ $77.89$} \vspace{1mm}\\
    
    \textbf{\trds{}} 
        & \makecell[c]{\lbtable \\ \gpttable} 
        & \makecell[c]{$90.8\faithDiff$ \\ $92.2\faithDiff$} 
        & \makecell[c]{$97.5\faithDiff$ \\ $98.1\faithDiff$} 
        & \makecell[c]{$92.2\faithDiff$ \\ $93.3\faithDiff$} 
        & \makecell[c]{$98.9\faithDiff$ \\ $98.1\faithDiff$} \\
    \bottomrule
    \multicolumn{6}{c}{ \textbf{\dls{}}} \\
    \toprule
    \textbf{RD} 
        & 
        & \makecell[c]{$40.7$} 
        & \makecell[c]{$56.3$} 
        & \makecell[c]{$51.0$} 
        & \makecell[c]{$70.7$} \\
    \midrule
    \textbf{\zs{}} 
        & \makecell[c]{\lbtable \\ \gpttable} 
        & \makecell[c]{$31.3\faithDiff$ \\ $46.2$} 
        & \makecell[c]{$50.6$ \\ $63.2$} 
        & \makecell[c]{$35.9\faithDiff$ \\ $56.1$} 
        & \makecell[c]{$70.0$ \\ $80.5\faithDiff$} \vspace{1mm}\\
    
    \textbf{\ods{}} 
        & \makecell[c]{\lbtable \\ \gpttable} 
        & \makecell[c]{$81.6\faithDiff$ \\ $93.8\faithDiff$} 
        & \makecell[c]{$92.2\faithDiff$ \\ $\mathbf{97.7}\faithDiff$} 
        & \makecell[c]{$85.3\faithDiff$ \\ $\mathbf{95.4}\faithDiff$} 
        & \makecell[c]{$96.9\faithDiff$ \\ $\mathbf{99.2}\faithDiff$} \vspace{1mm}\\
    
    \textbf{\tds{}} 
        & \makecell[c]{\lbtable \\ \gpttable} 
        & \makecell[c]{$81.4\faithDiff$ \\ $88.1\faithDiff$} 
        & \makecell[c]{$93.0\faithDiff$ \\ $95.6\faithDiff$} 
        & \makecell[c]{$83.1\faithDiff$ \\ $90.1\faithDiff$} 
        & \makecell[c]{$96.1\faithDiff$ \\ $96.8\faithDiff$} \vspace{1mm}\\

    \textbf{\tdsr{}} 
        & \makecell[c]{\lbtable \\ \gpttable} 
        & \makecell[c]{$56.0\faithDiff$ \\ $47.0$} 
        & \makecell[c]{$69.1\faithDiff$ \\ $57.0$} 
        & \makecell[c]{$64.7\faithDiff$ \\ $60.8\faithDiff$} 
        & \makecell[c]{$83.6\faithDiff$ \\ $75.7$} \vspace{1mm}\\
    
    \textbf{\trds{}} 
        & \makecell[c]{\lbtable \\ \gpttable} 
        & \makecell[c]{$84.4\faithDiff$ \\ $\mathbf{88.8}\faithDiff$} 
        & \makecell[c]{$93.8\faithDiff$ \\ $96.9\faithDiff$} 
        & \makecell[c]{$86.8\faithDiff$ \\ $90.8\faithDiff$} 
        & \makecell[c]{$96.7\faithDiff$ \\ $98.4\faithDiff$} \\
    \bottomrule
\end{tabular}
\end{scriptsize}

%% file: rag.tex
\section{Retrieval Augmented Generation (RAG)}
\label{sec:rag}
In the previous section we demonstrated the merits of corpus enrichment for ad hoc retrieval; specifically, we generated relevant documents which were retrievable at top ranks. We now turn to study the merits of applying corpus enrichment for question answering using LLMs. Specifically, we study the potential effectiveness improvement of retrieval augmented generation (RAG) \cite{zhao2024_rag_survey, ram2023_RALM, Huly_old_IR_rag} as a result of corpus enrichment. RAG has become a standard approach for improving Question-Answering (QA) performance by incorporating relevant external context into the generation process \cite{ram2023_RALM, lewis2020retrieval, izacard2021leveragingpassageretrievalgenerative}. In our QA task, given a question $\query$, a large language model (LLM), and a corpus $\corpus$, the objective is for the LLM to generate an answer to $\query$ using the information available in $\corpus$ and the information encoded in the parameters of the LLM.


We use the LLM to answer a set of questions with and without RAG. In the RAG setting, we first retrieve using \okapi{} five documents using the question as a query and then provide the LLM with the concatenated content of these documents as additional context to the question \cite{ram2023_RALM}; the prompt is presented in Figure \ref{rag_prompt}. To study the merits of corpus enrichment for the RAG task, we use two different corpora: the original corpus $\corpus$ and an enriched corpus $\corpuse$; recall that the enriched corpus depends on the generation method and the LLM used to generate a new document.

The setting without RAG serves as a baseline to the RAG setting: the LLM is prompted to answer a given question without being provided with any retrieved documents as context; the prompt is presented in Figure \ref{llm_prompt}.  

\subsection{Experimental setup}
\myparagraph{LLMs}  
We used \la{} and \lb{}\footnote{We use the same hyper-parameter settings reported in Section \ref{sec:direct}.} as our LLMs both for document generation and answer generation. \okapi{} BM25 was used to retrieve documents provided as context to the LLM to generate answers (in the RAG setting). The study of using alternative ranking functions is beyond the scope of this work.

\myparagraph{Datasets}
We used a version of the Natural Questions (NQ) dataset \cite{kwiatkowski-etal-2019-natural,lee_natural_questions_variant}, where the questions have short answers, limited to five words \cite{Bohnet+al:23a}. The dataset includes $3610$ natural questions issued to Google Search, with answers selected from the corresponding Wikipedia page.

We constructed the initial corpus using a dump of the English Wikipedia, dated December 20, 2018 \cite{karpukhin2020densepassageretrievalopendomain}. Each Wikipedia page was divided to disjoint passages of 100 words, henceforth referred to as documents \cite{karpukhin2020densepassageretrievalopendomain}. Each document included the title of the respective page along with its text; the title was concatenated to the document \cite{karpukhin2020densepassageretrievalopendomain,Huly_old_IR_rag}.

\myparagraph{Generating documents and answers}  
The prompts to the LLMs for answering questions without and with RAG are provided in Figures \ref{llm_prompt} and \ref{rag_prompt}. For corpus enrichment, we employed the generation methods described in Section \ref{sec:gen}, excluding the ZS method which does not use existing documents in the corpus. In Section \ref{sec:direct} we showed that its faithfulness was inferior to that of the generated documents produced by other generation methods.
We define a successful response as one in which the generated answer contains a correct answer, i.e., the string of the correct answer is included in the generated response.

\begin{figure}[t]
\centering
\input{Images/LLM}
\caption{The prompt to LLM with a question (\nrag{}).}
\label{llm_prompt}
\end{figure}

\begin{figure}[t]
\centering
\input{Images/RAG}
\caption{The prompt to LLM with a question and five documents as a context (RAG).}
\label{rag_prompt}
\end{figure}

\myparagraph{Document selection}  
Our generation methods utilize documents selected from the corpus. In Section \ref{sec:direct} we used known relevant documents. However, in the RAG setting we address here, no explicit relevance judgments are available. Instead, we used an oracle method for document selection. Therefore, as a proof of concept, we defined a document as relevant to a question if it contained the true answer. To prevent overlap between documents retrieved for RAG and those summarized to generate a new document, we selected the top five retrieved documents for RAG, while documents used for generation were selected from the positions 6 to 1000. We iterated over the ranked documents until we selected enough "relevant" documents for each generation method (up to three documents per question). For the \tdsr{} baseline generation method that is a summarization of a relevant and a randomly selected document from positions $6$ to $1000$, excluding documents that contained the answer. If an insufficient number of "relevant" documents were found for a given question and generation method, we did not generate a document for that question and method; for instance, in the \trds{} generation method (see Section \ref{sec:gen}), if a question had only two "relevant" documents available, we did not generate a new document for that question.
To ensure that documents in the corpus have approximately the same length, we truncated generated documents longer than $100$ words and discarded those shorter than $80$ words.

We generated a total of $19,359$ documents across all methods and LLMs. Specifically, $2,971$ and $2,683$ documents were generated by \gpt{} and \lb{}, respectively, for the \ods{} method; $2,708$ and $2,595$ for \tdsr{}; $2,191$ and $2,091$ for \tds{}; and $2,118$ and $2,002$ for \trds{}.


\subsection{Results}
In Table \ref{tab:rag} we present the comparison of several settings\footnote{We conducted the same statistical significance test as that in Section \ref{sec:direct}.}: (i) RAG with different enriched corpora\footnote{For each LLM and generation method.}, (ii) RAG without corpus enrichment, and (iii) \textbf{\nrag{}}: answers produced by the LLM without using retrieved documents. We report the following measures: (i) the percentage of correct answers (i.e., accuracy, denoted \textbf{Acc}), (ii) the percentage of questions in which a correct answer appeared in one of the top five documents retrieved by \okapi{} (denoted \textbf{Ans-5}), and (iii) the percentage of questions in which a
generated document was ranked in the top five documents (denoted
\textbf{Gen-5}). These measures are reported for all generation methods and for both LLMs (\la{} and \lb{}).

We can see in Table \ref{tab:rag} that for \la{}, applying RAG without corpus enrichment yields a higher accuracy of correct answers than generating an answer without RAG ($35.2\%$ vs. $33.5\%$). For \lb{}, generating answers without RAG results in higher accuracy than using RAG ($41.9\%$ vs. $38.3\%$) \cite{wu2025doesragintroduceunfairness}.

We can also see in Table \ref{tab:rag} that the enriched corpus yields statistically significant higher ratios of correct answers than the corpus without enrichment. For \lb{}, the accuracy ranges from $46.9\%$ (with \tdsr{}) to $49.3\%$ (with \trds{}). For \la{}, the accuracy ranges from $50.3\%$ (with \tdsr{}) to $57.3\%$ (with \ods{}). The increased accuracy with the enriched corpus may be attributed to the presence of a correct answer in at least one of the top five retrieved documents. Indeed, without enrichment, $45.9\%$ of the questions included a correct answer in the top five documents. However, with corpus enrichment, this percentage increases and ranges from $66.6\%$ to $77\%$ for \la{} and from $59.8\%$ to $61.4\%$ for \lb{}. Note that more than $40\%$ of the questions contained a generated document that was ranked among the top five documents provided to the RAG, which may have contributed to improved answer accuracy. Overall, \ods{} and \trds{} perform marginally better than the other generation methods for \la{} and \lb{}, respectively, with \la{} achieving higher performance across methods.




\begin{table}[t]
\caption{The effect of corpus enrichment on the effectiveness of RAG. We report: (i) the percentage of questions for which a correct answer was generated (\textbf{Acc}), (ii) the percentage of questions where the correct answer appears in one of the top five retrieved documents (\textbf{Ans-5}), and (iii) the percentage of questions where a generated document is ranked among the top five retrieved documents. These measures are reported for four generation methods and two LLMs (\la{} and \lb{}, denoted \textbf{L2} and \textbf{L3}, respectively). We report also the numbers for the corpus without enrichment serve as a baseline. We report the results both for generating answers without RAG (denoted \nrag{}) and for RAG without corpus enrichment. Statistically significant differences in \textbf{Acc} with the \nrag{} and RAG baselines are marked with "$l$" and "$r$", respectively. The best result in a column is boldfaced.}
\label{tab:rag}
\centering
\input{Tables/RAG_table}

\end{table}


%% file: Images/LLM.tex
\begin{lstlisting}[style=promptstyle]
PROMPT = "Instructions: Answer the question. Keep the answer concise. Question: <question>"
\end{lstlisting}

%% file: Images/RAG.tex
\begin{lstlisting}[style=promptstyle]
PROMPT = "Instructions: Answer the question based on the given passages below. Keep the answer concise. Passages: Passage 1: <passage #1> Passage 2: <passage #2> Passage 3: <passage #3> Passage 4: <passage #4> Passage 5: <passage #5> Question: <question>"
\end{lstlisting}

%% file: Tables/RAG_table.tex
\newcommand{\diffNORAG}{^{l}} 
\newcommand{\diffRAG}{_{r}} 
\begin{scriptsize}
\begin{tabular}{lcccc}
\toprule
 & & \makecell[c]{\textbf{Acc}} & \makecell[c]{\textbf{Ans-5}} & \makecell[c]{\textbf{Gen-5}} \\
\bottomrule
\multicolumn{5}{c}{\textbf{Baselines - without corpus enrichment}} \\
\toprule
\textbf{\nrag{}} & \makecell[c]{\latable \\ \lbtable} & \makecell[c]{$33.5$ \\ $41.9$} & \makecell[c]{$-$ \\ $-$} & \makecell[c]{$-$ \\ $-$} \\
\textbf{RAG} & \makecell[c]{\latable \\ \lbtable} & \makecell[c]{$35.2$ \\ $38.3$} & \makecell[c]{$45.9$ \\ $45.9$} & \makecell[c]{$-$ \\ $-$} \\
\bottomrule
\multicolumn{5}{c}{\textbf{RAG with corpus enrichment}} \\
\toprule
\textbf{\ods{}} & \makecell[c]{\latable \\ \lbtable} & \makecell[c]{$\mathbf{57.3}\diffNORAG\diffRAG$ \\ $49.2\diffNORAG\diffRAG$} & \makecell[c]{$\mathbf{77.0}$ \\ $61.4$} & \makecell[c]{$\mathbf{80.4}$ \\ $44.7$} \\
\textbf{\tds{}} & \makecell[c]{\latable \\ \lbtable} & \makecell[c]{$53.1\diffNORAG\diffRAG$ \\ $48.7\diffNORAG\diffRAG$} & \makecell[c]{$68.1$ \\ $60.3$} & \makecell[c]{$57.2$ \\ $41.4$} \\
\textbf{\tdsr{}} & \makecell[c]{\latable \\ \lbtable} & \makecell[c]{$50.3\diffNORAG\diffRAG$ \\ $46.9\diffNORAG\diffRAG$} & \makecell[c]{$67.1$ \\ $59.8$} & \makecell[c]{$59.1$ \\ $43.2$} \\
\textbf{\trds{}} & \makecell[c]{\latable \\ \lbtable} & \makecell[c]{$53.1\diffNORAG\diffRAG$ \\ $49.3\diffNORAG\diffRAG$} & \makecell[c]{$66.6$ \\ $59.9$} & \makecell[c]{$55.0$ \\ $41.2$} \\
\bottomrule
\end{tabular}
\end{scriptsize}

%% file: attribution.tex
\section{ATTRIBUTION}
\label{sec:att}
In this section, we discuss the potential of using corpus enrichment to improve attribution in QA settings. Attribution refers to presenting a passage that supports the correctness of a given answer \cite{Bohnet+al:23a}. Specifically, inspired by Bohnet et al. \cite{Bohnet+al:23a}, who proposed a methodology to estimate whether a document from a corpus attributes an answer, we evaluate the attribution effectiveness in the RAG setting discussed in Section \ref{sec:rag}; i.e., we use RAG to generate answers.

\subsection{Experimental setting}
An attributed QA task is defined as follows: Given a question $\query$ and a generated answer $\answer$, we aim to find a document\footnote{In practice, these are passages of documents.} from a corpus that attributes the answer $\answer$; that is, provides support to its correctness for $\query$.
We focus on an evaluation approach for attribution which utilizes the NLI-based method proposed by Bohnet et al. \cite{Bohnet+al:23a}. This method estimates the level of attribution of a candidate document $d$ to an answer $\answer$. Specifically, given an answer $\answer$, the model computes the extent to which $d$ entails $\answer$ in response to the question $\query$. The resulting NLI score is from [0,1], where a score above 0.5 indicates an entailment (i.e., $d$ is considered to attribute $\answer$).
Bohnet et al. \cite{Bohnet+al:23a} demonstrated that the estimates assigned by their approach correlate with the human-annotated attribution scores introduced by Rashkin et al. \cite{rashkin_measure_attribution}.

\myparagraph{Candidate document for attribution}
We employ two methods to retrieve a document $\datt$ as a candidate to attribute an answer $\answer$ to a question $\query$. In both methods, we begin by ranking the documents in the corpus using the \okapi{} BM25 retrieval method. In the first method (denoted \textbf{\okr{}}), we select the top-ranked document as a candidate for attribution. In the second method (denoted \textbf{\oknr{}}) \cite{Bohnet+al:23a} we first retrieve the top 50 documents using \okapi{}, and then re-rank them based on their NLI scores with respect to $\answer$. The top-ranked document is selected.
For both method we then compute the percentage of questions where the selected document entails the answer (i.e., the \textbf{entailment accuracy}).

In cases where the candidate document is a document generated by one of our methods, we also report a modified version of the entailment accuracy, which considers for attribution only human-authored documents from the original corpus $\corpus$; the modified entailment accuracy is the highest entailment accuracy determined by one of the documents which served for generation. Recall that every generated document\footnote{Except the zero-shot (\zs{}) generation method which is not used in Sections \ref{sec:rag} and \ref{sec:att}.} was generated using one or more existing documents from the original corpus $\corpus$.

\myparagraph{Experimental setting}
Our attribution setup closely follows the RAG setup from Section \ref{sec:rag}, using the same datasets and models for document and answer generation. To analyze the impact of corpus enrichment on attribution, we compare four settings, each used RAG to generate answers: (i) generating answers and selecting attribution documents using the original, non-enriched corpus $\corpus$, (ii) generating answers using an enriched corpus $\corpuse$ but selecting attribution documents using the original corpus $\corpus$, (iii) generating answers using the original corpus $\corpus$ while selecting attribution documents using the enriched corpus $\corpuse$, and (iv) generating answers and selecting attribution documents using the enriched corpus $\corpuse$. The goal of using these settings is to evaluate the merits of corpus enrichment in two capacities: used for RAG and for attribution (and both). The first two settings, where the original corpus is used for attribution, serve as baselines for the other two settings that utilize the enriched corpus for attribution.

\subsection{Results}  
Table \ref{tab:att} presents the attribution evaluation numbers\footnote{We used the same statistical significance test as that in Section \ref{sec:rag}.}. We report: (i) the percentage of questions for which the candidate document contained the answer (denoted \textbf{CA}), (ii) the entailment accuracy (denoted \textbf{\aisp{}}), and (iii) the entailment accuracy using documents only from the original corpus $\corpus$ for entailment (denoted \textbf{\maisp{}}).

We can see in Table \ref{tab:att} that in all the settings, the levels of \maisp{} are consistently lower than \aisp{}; recall that \maisp{} was computed solely based on documents from the original corpus while \aisp{} was computed from the enriched corpus. This finding attests to the merits of using corpus enrichment for attribution.

We see in Table \ref{tab:att} that among the two baseline settings (attribution without corpus enrichment) the \aisp{} for RAG without corpus enrichment is marginally higher than that with enrichment for \okr{}. For the \oknr{} ranker we observe an opposite trend: the \aisp{} for RAG without corpus enrichment is lower than that of RAG using corpus enrichment (excluding \tdsr{}). These findings suggest that when the attribution documents are not derived from an enriched corpus, both settings yield similar performance. In the setting in which RAG was enriched, the LLM generates answers in many cases based on the newly generated document. Yet, this document is not available for attribution by the setting definition. Thus, attribution is based on other documents which might not include the answer, or information pertaining to it. We note that the reason to analyze this setting is to decouple the corpus enrichment for RAG from the corpus enrichment for attribution and to study their effectiveness separately. Below we analyze the setting where both RAG and attribution are performed over an enriched corpus.



We also see in Table \ref{tab:att} that among the two settings of RAG without corpus enrichment, the setting in which attribution is using corpus enrichment the \aisp{} is slightly higher than that of the setting in which attribution is without corpus enrichment. This finding can be attributed to the selection of the candidate document for attribution: without corpus enrichment, the percentage of questions for which the top candidate document contained the answer was lower than in the setting with corpus enrichment for attribution. Notably, when enriching the corpus for attribution, a larger number of documents exist that can potentially support the generated answer.

Table \ref{tab:att} also shows that in the setting we enriched both the corpus used for RAG and the corpus used for attribution resulted in higher levels of \aisp{} and \maisp{} than the other three settings, for both ranking methods (\okr{} and \oknr{}).

These results highlight the importance of enriching the corpus not only for RAG but also for attribution in order to achieve the higher levels of attribution levels. The highest observed \aisp{} value was $74.5\%$, obtained when both RAG and attribution were with corpus enrichment, using \la{} with the \ods{} generation method for \oknr{}. The highest observed \maisp{} value is $48.6\%$, obtained using \lb{} with the \trds{} generation method for \oknr{}.


\begin{table}[t]
\caption{Attributed QA with corpus enrichment. We report the following measures: (i) the percentage of questions for which the candidate document (i.e., the top-ranked document) contains the answer (\textbf{CA}), (ii) the entailment accuracy (\aisp{}), and (iii) the entailment accuracy based on the original corpus (\maisp{}). We compare four settings; RAG without and with corpus enrichment are denoted \ragne{} and \rage{}. These measures are reported for four settings, across four generation methods, two LLMs (\la{} and \lb{}, denoted \textbf{L2} and \textbf{L3}), and two ranking methods used to select the candidate document for attribution (\okr{} and \oknr{}). Statistically significant differences in the measures (except for CA) are marked with "$*$". The best result in a column is boldfaced.}
\label{tab:att}
\input{Tables/ATT_table}
\end{table}

%% file: Tables/ATT_table.tex
\newcommand{\diffRAGNE}{^{*}} 
\newcommand{\diffRAGE}{^{*}} 
\begin{scriptsize}
\setlength{\tabcolsep}{3pt}
\begin{tabular}{lllcccccc}
\toprule
\multicolumn{3}{l}{} 
& \multicolumn{3}{c}{\textbf{\okr{}}} 
& \multicolumn{3}{c}{\textbf{\oknr{}}} \\
\cmidrule(lr){4-6} \cmidrule(lr){7-9}
& & & \textbf{CA} & $\mathbf{\aisp{}}$ & $\mathbf{\maisp{}}$
  & \textbf{CA} & $\mathbf{\aisp{}}$ & $\mathbf{\maisp{}}$ \\
\toprule
\multicolumn{9}{c}{\textbf{Baselines - Attribution without corpus enrichment}} \\
\toprule

\textbf{\ragne{}} & & \makecell[c]{\latable \\ \lbtable} & \makecell[c]{$ 22.1 $ \\ $ 22.1 $} & \makecell[c]{$ 14.7 $ \\ $ 18.0 $} & \makecell[c]{$ - $ \\ $ - $} 
  & \makecell[c]{$ 36.4 $ \\ $ 32.3 $} & \makecell[c]{$ 43.0 $ \\ $ 46.0 $} & \makecell[c]{$ - $ \\ $ - $} \\
\midrule

\multirow{8}{*}{\textbf{\rage{}}} 
& \textbf{\ods{}} & \makecell[c]{\latable \\ \lbtable} & \makecell[c]{$ 22.1 $ \\ $ 22.1 $} & \makecell[c]{$ 9.8 $ \\ $ 14.4 $} & \makecell[c]{$ - $ \\ $ - $} 
  & \makecell[c]{$ 54.1 $ \\ $ 40.4 $} & \makecell[c]{$ 43.1 $ \\ $ 46.4 $} & \makecell[c]{$ - $ \\ $ - $} \\

& \textbf{\tds{}} & \makecell[c]{\latable \\ \lbtable} & \makecell[c]{$ 22.1 $ \\ $ 22.1 $} & \makecell[c]{$ 10.9 $ \\ $ 15.3 $} & \makecell[c]{$ - $ \\ $ - $} 
  & \makecell[c]{$ 53.5 $ \\ $ 41.1 $} & \makecell[c]{$ 45.0 $ \\ $ 48.5 $} & \makecell[c]{$ - $ \\ $ - $} \\

& \textbf{\tdsr{}} & \makecell[c]{\latable \\ \lbtable} & \makecell[c]{$ 22.1 $ \\ $ 22.1 $} & \makecell[c]{$ 10.1 $ \\ $ 14.8 $} & \makecell[c]{$ - $ \\ $ - $} 
  & \makecell[c]{$ 52.9 $ \\ $ 40.9 $} & \makecell[c]{$ 42.9 $ \\ $ 46.9 $} & \makecell[c]{$ - $ \\ $ - $} \\

& \textbf{\trds{}} & \makecell[c]{\latable \\ \lbtable} & \makecell[c]{$ 22.1 $ \\ $ 22.1 $} & \makecell[c]{$ 11.2 $ \\ $ 15.5 $} & \makecell[c]{$ - $ \\ $ - $} 
  & \makecell[c]{$ 52.9 $ \\ $ 41.6 $} & \makecell[c]{$ 45.0 $ \\ $ 48.9 $} & \makecell[c]{$ - $ \\ $ - $} \\
\toprule
\multicolumn{9}{c}{\textbf{Attribution with corpus enrichment}} \\
\toprule

\multirow{8}{*}{\textbf{\ragne{}}} 
& \textbf{\ods{}} & \makecell[c]{\latable \\ \lbtable} & \makecell[c]{$\mathbf{66.7}$ \\ $ 39.3 $} & \makecell[c]{$ 20.6\diffRAGNE $ \\ $ 17.6 $} & \makecell[c]{$ 11.4\diffRAGNE $ \\ $ 15.8\diffRAGNE $} 
  & \makecell[c]{$ 39.9 $ \\ $ 34.7 $} & \makecell[c]{$ 47.6\diffRAGNE $ \\ $ 47.2 $} & \makecell[c]{$ 39.1\diffRAGNE $ \\ $ 44.4 $} \\

& \textbf{\tds{}} & \makecell[c]{\latable \\ \lbtable} & \makecell[c]{$ 54.5 $ \\ $ 40.1 $} & \makecell[c]{$ 18.8\diffRAGNE $ \\ $ 18.2 $} & \makecell[c]{$ 15.4 $ \\ $ 17.4 $} 
  & \makecell[c]{$ 39.0 $ \\ $ 34.9 $} & \makecell[c]{$ 45.8\diffRAGNE $ \\ $ 47.6 $} & \makecell[c]{$ 40.9 $ \\ $ 45.0 $} \\

& \textbf{\tdsr{}} & \makecell[c]{\latable \\ \lbtable} & \makecell[c]{$ 51.8 $ \\ $ 38.9 $} & \makecell[c]{$ 18.2\diffRAGNE $ \\ $ 16.9 $} & \makecell[c]{$ 13.2 $ \\ $ 16.0\diffRAGNE $} 
  & \makecell[c]{$ 39.0 $ \\ $ 34.6 $} & \makecell[c]{$ 45.6\diffRAGNE $ \\ $ 47.6 $} & \makecell[c]{$ 40.1\diffRAGNE $ \\ $ 44.9 $} \\

& \textbf{\trds{}} & \makecell[c]{\latable \\ \lbtable} & \makecell[c]{$ 54.1 $ \\ $ 40.6 $} & \makecell[c]{$ 20.3\diffRAGNE $ \\ $ 18.9 $} & \makecell[c]{$ 16.9\diffRAGNE $ \\ $ 18.5 $} 
  & \makecell[c]{$ 39.1 $ \\ $ 34.9 $} & \makecell[c]{$ 46.2\diffRAGNE $ \\ $ 47.6 $} & \makecell[c]{$ 41.4 $ \\ $ 45.1 $} \\

\midrule

\multirow{8}{*}{\textbf{\rage{}}} 
& \textbf{\ods{}} & \makecell[c]{\latable \\ \lbtable} & \makecell[c]{$\mathbf{66.7}$ \\ $ 39.3 $} & \makecell[c]{$\mathbf{61.4}\diffRAGE $ \\ $ 30.1\diffRAGE $} & \makecell[c]{$ 26.3\diffRAGE $ \\ $ 21.3\diffRAGE $} 
  & \makecell[c]{$\mathbf{65.7}$ \\ $ 47.7 $} & \makecell[c]{$\mathbf{74.5}\diffRAGE $ \\ $ 56.1\diffRAGE $} & \makecell[c]{$ 39.6\diffRAGE $ \\ $ 44.4 $} \\

& \textbf{\tds{}} & \makecell[c]{\latable \\ \lbtable} & \makecell[c]{$ 54.5 $ \\ $ 40.1 $} & \makecell[c]{$ 46.5\diffRAGE $ \\ $ 30.0\diffRAGE $} & \makecell[c]{$ 26.9\diffRAGE $ \\ $ 23.6\diffRAGE $} 
  & \makecell[c]{$ 61.8 $ \\ $ 48.0 $} & \makecell[c]{$ 65.8\diffRAGE $ \\ $ 56.0\diffRAGE $} & \makecell[c]{$ 44.1 $ \\ $ 47.5 $} \\

  & \textbf{\tdsr{}} & \makecell[c]{\latable \\ \lbtable} & \makecell[c]{$ 51.8 $ \\ $ 38.9 $} & \makecell[c]{$ 46.3\diffRAGE $ \\ $ 28.4\diffRAGE $} & \makecell[c]{$ 23.4\diffRAGE $ \\ $ 21.2\diffRAGE $} 
  & \makecell[c]{$ 60.2 $ \\ $ 47.3 $} & \makecell[c]{$ 65.8\diffRAGE $ \\ $ 54.4\diffRAGE $} & \makecell[c]{$ 41.0 $ \\ $ 45.3 $} \\

& \textbf{\trds{}} & \makecell[c]{\latable \\ \lbtable} & \makecell[c]{$ 54.1 $ \\ $ 40.6 $} & \makecell[c]{$ 48.0\diffRAGE $ \\ $ 31.2\diffRAGE $} & \makecell[c]{$ \mathbf{28.4}\diffRAGE $ \\ $ 25.0\diffRAGE $} 
  & \makecell[c]{$ 61.1 $ \\ $ 48.5 $} & \makecell[c]{$ 66.2\diffRAGE $ \\ $ 56.5\diffRAGE $} & \makecell[c]{$ 45.1 $ \\ $\mathbf{48.6}$} \\

\bottomrule
\end{tabular}
\end{scriptsize}

%% file: conc.tex
\section{Conclusion and Future Work}

In this paper we argued for a novel perspective of using genAI technology to enrich a corpus with the goal of improving the effectiveness of query-based retrieval. That is, using large language models (LLMs), existing documents are modified and/or new documents are generated, to improve retrieval effectiveness. Thus, while LLMs are often used for direct response generation, to generate IDs of documents relevant to a query \cite{li2025matchinggenerationsurveygenerative}, or to modify queries \cite{Jagerman+al:23a}, we proposed to use them to enrich a corpus.

As an initial study intended to support the merits of corpus enrichment for query-based retrieval, we presented a few simple prompting-based methods for document modification and generation. We showed that enriching the corpus with the generated documents can help to improve the effectiveness of standard ad hoc retrieval. Furthermore, we showed how corpus enrichment can help to improve the performance of question answering using LLMs when applying RAG (retrieval augmented generation). Additional application for which we demonstrated the merits of corpus enrichment was attribution in question answering: retrieving passages that provide support to the correctness of an answer.

An important aspect of enriching a corpus using newly generated documents is the faithfulness of the information they contain to that in the corpus. We proposed measures for quantifying this faithfulness.

Our evaluation of RAG with corpus enrichment was based on an oracle experiment where we used documents (passages) that contain the answer to generate new documents. For future work we plan to devise methods that select documents to serve as the basis for generation; e.g., those most similar to the question in some embedding space. In a related vein, using pseudo relevant documents for generation of new documents in the standard ad hoc retrieval setting is a direction we intend to explore.

We also plan to study the merits of using corpus expansion to reduce the retrieval performance variance among queries representing the same information need \cite{Buckley+al:94a, Carpineto:Romano:12a}. We will address this challenge for specific retrieval functions and in a way that is agnostic to the retrieval function.

Increasing the likelihood that a generated document is relevant, e.g., by using improved prompting techniques and/or more advanced methods to select the documents based on which generation is performed, is an additional future direction we intend to pursue.